\documentclass{aa}
\usepackage{graphicx}
\usepackage{amssymb}
\usepackage{amsmath}            

\begin{document}

\title{Desorption of CO and O$_2$ interstellar ice analogs}
\author{ K. Acharyya \inst{1,2}, G.W. Fuchs \inst{1}, H.J. Fraser
\inst{3}, E.F. van Dishoeck \inst{1}, and H. Linnartz \inst{1} }
\institute{ Raymond and Beverly Sackler Laboratory for Astrophysics,
Leiden Observatory, \\ Leiden University, Postbus 9513, NL 2300 RA Leiden, The
Netherlands \\ 
\email{acharyya@strw.leidenuniv.nl} \and Centre For
Space Physics, 43 Chalantika, Garia, Kolkata, 720084, India \and
Department of Physics SUPA (Scottish Universities Physics Alliance), University of Strathclyde, 107 Rottenrow East,
Glasgow G4 ONG, Scotland } 
\date{07/08/2006} 

\abstract{ } { Solid O$_2$ has been proposed as a possible reservoir
for oxygen in dense clouds through freeze-out processes.  The aim of
this work is to characterize quantitatively the physical processes
that are involved in the desorption kinetics of CO-O$_2$ ices by
interpreting laboratory temperature programmed desorption (TPD) data.
This information is used to simulate the behavior of CO-O$_2$ ices
under astrophysical conditions.  }  {The TPD spectra have been
recorded under ultra high vacuum conditions for pure, layered and
mixed morphologies for different thicknesses, temperatures and mixing
ratios. An empirical kinetic model is used to interpret the results
and to provide input parameters for astrophysical models.  }  {Binding
energies are determined for different ice morphologies. Independent of
the ice morphology, the desorption of O$_2$ is found to follow
0$^{th}$-order kinetics.  Binding energies and temperature-dependent
sticking probabilities for CO--CO, O$_2$--O$_2$ and CO--O$_2$ are
determined. O$_2$ is slightly less volatile than CO, with binding
energies of 912$\pm$15 versus 858$\pm$15 K for pure ices. In mixed and
layered ices, CO does not co-desorb with O$_2$ but its binding
energies are slightly increased compared with pure ice whereas those
for O$_2$ are slightly decreased.  Lower limits to the sticking
probabilities of CO and O$_2$ are 0.9 and 0.85, respectively, at
temperatures below 20~K.  The balance between accretion and
desorption is studied for O$_2$ and CO in astrophysically relevant
scenarios. Only minor differences are found between the two species,
i.e., both desorb between 16 and 18~K in typical environments around
young stars.  Thus, clouds with significant abundances of gaseous CO
are unlikely to have large amounts of solid O$_2$.} { }
\titlerunning{Laboratory studies of CO and O$_2$ desorption}
\authorrunning{Acharyya et al.}


\maketitle
                
\section{Introduction}

Gas-grain interactions play a key role in the chemical evolution of
star-forming regions.  During the first stages of star formation
virtually all species accrete onto grains in dense cold cores. Later
on in the star formation sequence --- when so-called hot cores are
formed --- grains are warmed to temperatures where molecules can desorb
again. In order to characterize this astrophysical process
quantitatively it is necessary to understand the underlying molecular
physics by studying interstellar ice analogs under laboratory
controlled conditions. A series of recent papers shows that even for
simple molecules such a quantification is far from trivial
(e.g. Fraser et al. 2001, Collings et al. 2003, 2004, {\"O}berg et
al. 2005, Bisschop et al. 2006).  In the present work results for CO
and O$_2$ ices are discussed that extend recent work comparing
CO-N$_2$ and CO-O$_2$ ice features (Fuchs et al. 2006) to an empirical
kinetic model characterizing the desorption behavior.  

The reason for focusing on ices containing O$_2$ is that a substantial
amount of interstellar oxygen may well freeze out onto grains in the
form of molecular oxygen.  Attempts to determine gaseous
O$_2$-abundances from recent SWAS and ODIN campaigns put upper limits
on the O$_2$ abundance in cold dark clouds in the range of
3$\times$10$^{-6}$-1$\times$10$^{-7}$ (Goldsmith et al. 2000, Pagani
et al. 2003, Liseau et al. 2005). This low abundance, along with the
low abundance of gaseous H$_2$O, raises serious questions about the
total oxygen budget when compared with the well observed atomic oxygen
abundance of 3$\times 10^{-4}$ in diffuse clouds (Meyer et al. 1998,
Ehrenfreund \& van Dishoeck 1998).  One possible explanation for the
`missing' oxygen is that it is frozen out onto grains in the coldest
regions.  The fundamental vibration of solid O$_2$ around 1550
cm$^{-1}$ (6.45 $\mu$m) becomes observable through perturbations of
the symmetry of O$_2$ in a matrix or ice containing other molecules
(Ehrenfreund et al. 1992). This band has been sought towards the
proto-stellar sources RCrA IRS2 and NGC 7538 IRS9 (Vandenbussche et
al.\ 1999), but not detected. Upper limits between 50\% and 100\% of
solid O$_2$ relative to solid CO have been reported from analysis of
the CO profile, since solid CO, in contrast with solid O$_2$, has been
observed through its vibrational band at 4.67 $\mu$m (2140 cm$^{-1}$)
(e.g. Chiar et al. 1998, Pontoppidan et al. 2003). Here the amount of
frozen O$_2$ can be estimated by observing its influence on the shape
of the CO absorption band.  Transmission spectra recorded for mixed
CO-O$_2$ ices show indeed significant changes compared to pure CO ices
(Ehrenfreund et al. 1997, Elsila et al. 1997).  However, such changes
can also be caused by grain size and shape effects (Dartois 2006).
The best limits on solid O$_2$ therefore come
from analysis of the weak solid $^{13}$CO band which is not affected
by grain shape effects. This band leads to upper limits of 100\% on
the amount of O$_2$ that can be mixed with CO (Boogert et al.\ 2003,
Pontoppidan et al.\ 2003).

\begin{figure}[t]
   \includegraphics[width=8cm]{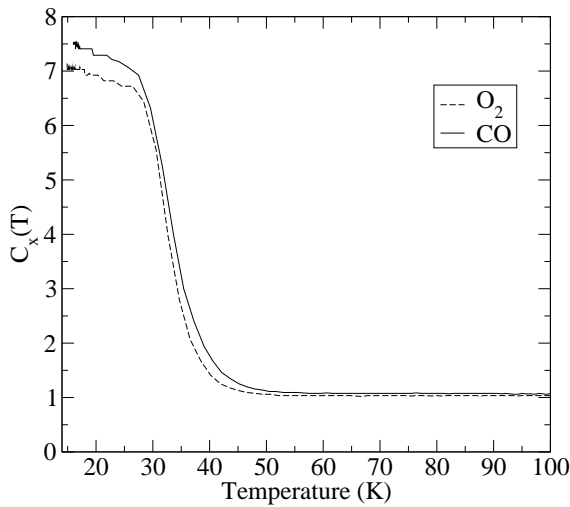}
   \vspace{0.0cm}
 \caption{The cryo-pumping factor ($C_x(T)= p_{rt}/p_{t}$) as a function of
           temperature for CO (solid) and O$_2$ (dashed).} 
\end{figure}

In this study, laboratory results for CO-O$_2$ ices are presented,
both in pure, layered and mixed ice morphologies for varying ice
thicknesses and for different relative abundances (Section 2). The
focus is on temperature programmed desorption (TPD) data that are
recorded to study the desorption process and to visualize changes in
ice morphology during heating (Section 3).  An empirical kinetic model
is used to interpret the TPD data (Section 4). The aim of this work is
to derive accurate molecular parameters (CO-CO, CO-O$_2$ and
O$_2$-O$_2$ binding energies, classification of desorption kinetics,
and temperature dependent sticking coefficients) to allow reliable
predictions of typical behavior under astrophysical conditions
(Section 5).

\section{Experiment}
\label{exp-sec}
The experimental setup consists of an ultra-high vacuum setup
(background pressure $<$ 5$\times$10$^{-11}$ mbar) in which
interstellar ice analogs are grown with mono-layer precision on a
2.5$\times$2.5 cm$^2$ sample surface made out of a 0.1 $\mu$m thick
polycrystalline gold film. Physical-chemical interactions can be
studied using TPD and/or reflection absorption infrared spectroscopy
(RAIRS).  Details of the setup are available from van
Broekhuizen (2005) and Fuchs et al. (2006) and here only relevant
details are given. We assume that for a sticking coefficient of 1,
molecules hitting the cold surface (14 K) build up one monolayer (ML)
at a steady gas exposure of 1.33$\cdot$10$^{-6}$ mbar s$^{-1}$
(10$^{-6}$ Torr s$^{-1}$). This is by definition one Langmuir [L] and
corresponds to roughly 10$^{15}$ particles cm$^{-2}$ in a non-porous
condensed solid. In the present experiment amorphous ice is grown
using a 0.01 L s$^{-1}$ growth velocity resulting in a well specified
layer thickness as long as the growth is linear with the exposure
time. Typical layer thicknesses are generated in the range between 20
and 80~L as astrophysical observations of young stellar objects in
nearby low-mass star-forming clouds indicate that CO ices exist at
thicknesses around 40 monolayers (Pontoppidan et al. 2003).  In
addition, astrochemical models suggest that O$_2$ may freeze out onto
a pre-existing CO layer and consequently both pure, mixed and layered
ices are studied here (Hasegawa et al. 1992, d'Hendecourt et al. 1985,
Bergin et al. 2000). Studies of O$_2$ in an H$_2$O-rich environment
have been performed by Collings et al.\ (2004).
  \begin{table}[t]
    \begin{center}
   \caption[]{Overview of all ice morphologies used in the CO-O$_2$ experiments. The 
    thickness is given in Langmuir [L].}
    \begin{tabular}{|l|c|c|c|c|}
    \hline
    \hline
        &  $^{13}$CO  \hspace{+0.25cm} $^{18}$O$_2$    & Total    &    $^{12}$CO \hspace{+0.25cm} $^{16}$O$_2$  & Total \\
    \hline
CO      &  20     \hspace{0.25cm}           -            &  20      &     20     \hspace {0.25cm}        -             & 20    \\
pure    &  40     \hspace{0.25cm}           -            &  40      &     40     \hspace {0.25cm}        -             & 40    \\
        &  60     \hspace{0.25cm}           -            &  60      &     -      \hspace {0.25cm}        -             & -     \\
        &  80     \hspace{0.25cm}           -            &  80      &     80     
\hspace {0.25cm}        -             & 80    \\
    \hline
O$_2$&      -   \hspace{0.25cm}            20            &  20      &     -      
\hspace {0.25cm}        20            & 20    \\
pure    &   -   \hspace{0.25cm}            40            &  40      &     -      
\hspace {0.25cm}        40            & 40    \\
        &   -   \hspace{0.25cm}            60            &  60      &     -      
\hspace {0.25cm}        -             & -     \\
        &   -   \hspace{0.25cm}            80            &  80      &     -      
\hspace {0.25cm}        80            & 80    \\
    \hline
CO:O$_2$&  20   \hspace{0.25cm}            20            &  40      &     -      \hspace{0.25cm}         -             & -     \\
mixed   &  30   \hspace{0.25cm}            30            &  60      &     -      \hspace{0.25cm}         -             & -     \\ 
        &  40   \hspace{0.25cm}            40            &  80      &     40     \hspace{0.25cm}         40            & 80    \\
        &  60   \hspace{0.25cm}            60            &  120     &     -      \hspace{0.25cm}         -             & -     \\
        &  80   \hspace{0.25cm}            80            &  160     &     -      \hspace{0.25cm}         -             & -     \\
    \hline
O$_2$/CO $^a$&  20   \hspace{0.25cm}            20            &  40      &     20     
\hspace{0.25cm}         20            & 40    \\
layered &  30   \hspace{0.25cm}            30            &  60      &     -      
\hspace{0.25cm}         -             & 80    \\
        &  40   \hspace{0.25cm}            40            &  80      &     40     
\hspace{0.25cm}         40            & 80    \\
        &  60   \hspace{0.25cm}            60            &  120     &     -      
\hspace{0.25cm}         -             & -     \\
        &  80   \hspace{0.25cm}            80            &  160     &     80     
\hspace{0.25cm}         80            & -     \\
        &  40   \hspace{0.25cm}         \,\, 5            &  45      &     -      
\hspace{0.25cm}         -             & -     \\
        &  40   \hspace{0.25cm}            10            &  50      &     40     
\hspace{0.25cm}         10            & 50    \\
        &  40   \hspace{0.25cm}            20            &  60      &     40     
\hspace{0.25cm}         20            & 60    \\
        &  40   \hspace{0.25cm}            30            &  70      &     -      
\hspace{0.25cm}         -             & -     \\
    \hline
CO/O$_2$ $^a$&  20   \hspace{0.25cm}            20            &  40      &     -      
\hspace{0.25cm}         -             & 40    \\
layered &  40   \hspace{0.25cm}            40            &  80      &     40     
\hspace{0.25cm}         40            & 80    \\
        &  80   \hspace{0.25cm}            80            &  160     &     -      
\hspace{0.25cm}         -             & -     \\
    \hline
    \hline
    \end{tabular}
$^a$ See text for explanation of the used notation.
    \label{table-1}
  \end{center}
   \end{table}

Two sets of ices have been investigated, $^{13}$CO-$^{18}$O$_2$ and
$^{12}$CO-$^{16}$O$_2$, both using isotopes of at least 99\%
purity. Most experiments were performed using $^{13}$CO-$^{18}$O$_2$
to distinguish between possible impurities in the vacuum system.  An
overview of the used ice samples is given in Table 1. Throughout this
paper, the notation x/y and x:y denote layered (x on top of y) and
mixed ice morphologies, respectively. The notation 1/1 indicates
layered ices of equal thickness, 1:1 refers to equally mixed ices and
x/40 L stands for ices where the thickness of the top layer is varied
by x~L and the bottom layer is kept constant at 40 L.

\begin{figure}[t]
\centering
    \includegraphics[angle=270,width=9cm]{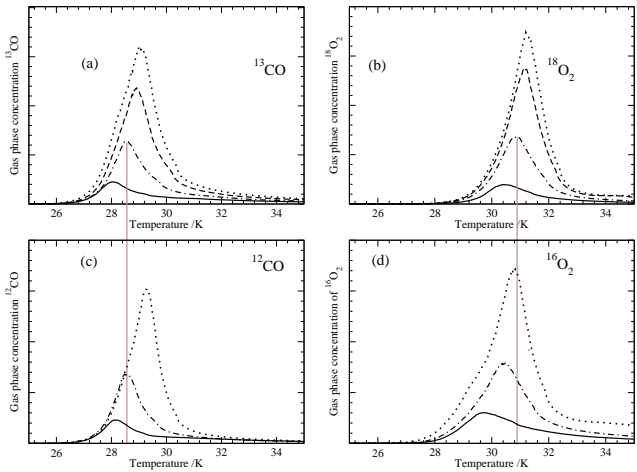}
    \caption{TPD spectra of pure ices. (a) $^{13}$CO and (b) $^{18}$O$_2$ - for
    exposures of 20 L (solid), 40 L (dot-dashed), 
    60 L (dashed) and 80 L (dotted) - and (c) $^{12}$CO and (d) $^{16}$O$_2$ 
    - for exposures of 20 L (solid), 40 L (dot-dashed) 
    and 80 L (dotted). The grey vertical lines indicate the peak temperatures at 40 L.}
\end{figure}

Once an ice is grown the desorption behavior is examined by a
controlled linear temperature rise from 12 to 80 K. The heating rate
in all experiments is 0.1 K min$^{-1}$, unless stated differently. The
CO and O$_2$ molecules that are released are monitored by a quadrupole
mass spectrometer (QMS). For a constant pumping speed the QMS signal
strengths are proportional to the amount of desorbed species. Since
the deposition surface, together with other cold surface areas like the radiation shield, 
can be regarded as a cryogenic pump a
temperature dependent cryo-pump factor has to be introduced as the
pressure in the chamber decreases when the temperature drops below 50
K. This effect is measured by opening the leakage valve up to a
certain pressure ($p_{rt}$) at room temperature after which the system
is cooled down to 14~K and the corresponding pressure ($p_t$) is
determined.  In Fig. 1 the ratio $C_x(T)= p_{rt}/p_{t}$ is shown as a
function of temperature for both CO and O$_2$ revealing that at 14 K
the cryo-pumping factor increases by a factor of about 7 with respect
to the pumping speed, $s_{tp}$, of the turbo pump alone.  This effect
is important and will be explicitly taken into account in the kinetic
model.

Reflection-Absorption Infrared (RAIR) spectra have been taken
simultaneously with the TPD data but are not included here. Spectra
for selected CO-O$_2$ experiments were presented and discussed in Fuchs
et al.\ (2006); they provide additional information on the processes
occurring in the ice during heating.

\section{Experimental results}
\label{exp-res-sec}
TPD spectra of pure and mixed/layered CO-O$_2$ ices are shown in
Figs. 2 and 3, respectively.

\vspace{2mm} \noindent {\bf Pure ices.}  In the case of pure $^{13}$CO
ice (Fig. 2a) and pure $^{12}$CO ice (Fig. 2c) the desorption starts
around 26.5 K and peaks between 28 and 29 K. All leading edges clearly
coincide and peak temperatures shift towards higher temperatures for
higher thicknesses. This behavior is typical for a 0$^{th}$-order
desorption process (see \S 4.1) and corresponds to a desorption rate
that is independent of the ice thickness and that remains constant
until no molecules are left on the surface (see also Collings et
al. 2003, Bisschop et al. 2006). The binding energies for both
isotopes are similar and consequently nearly the same peak temperature
is found.  For pure $^{18}$O$_2$ ice (Fig. 2b) and pure
$^{16}$O$_2$ ice (Fig. 2d) a similar behavior is found as for CO: all
desorption curves have the same leading edges and the peak temperature
slightly shifts for increasing thicknesses (from 29.5 to 31.3~K).  The
desorption peak of O$_2$ is clearly shifted to higher temperatures by
$\sim$2~K compared with CO.  A small shift of peak temperatures of
less than 0.5~K is observed when comparing $^{18}$O$_2$ and
$^{16}$O$_2$.

\begin{figure}[t]
    \includegraphics[angle=270,width=9cm]{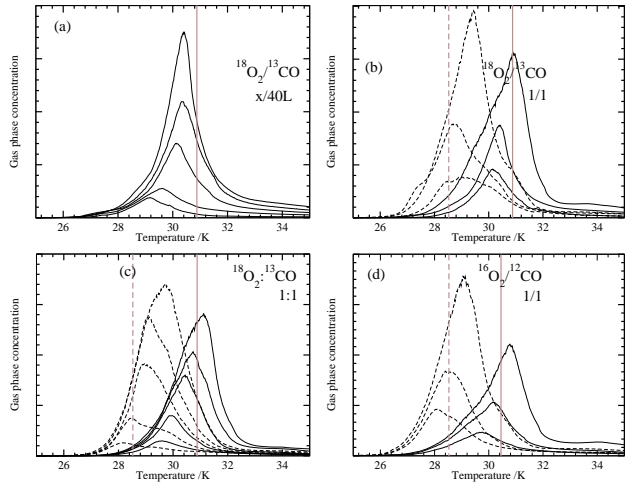}
    \caption{
     TPD spectra for mixed/layered ices. (a) O$_2$ trace of (5, 10, 20, 30 and 40 L)
     $^{18}$O$_2$/40 L $^{13}$CO - differential layer, 
    (b) (20, 40 and 80 L) $^{18}$O$_2$ (solid line) and $^{13}$CO (dashed
     line) for 1/1 O$_2$/CO system. 
    (The shoulder observed in the 40 L measurement is an experimental artifact
absent in other experiments).
    (c) (10, 20, 40, 60, 80 L) $^{18}$O$_2$ 
    (solid line) and $^{13}$CO (dashed line) for 1:1 mixed ices, (d) (20, 40 and 80
    L) $^{16}$O$_2$ (solid line) and $^{12}$CO 
    (dashed line) for the 1/1 O$_2$/CO system. 
    The grey vertical lines indicate the peak temperatures for pure O$_2$ (solid) and pure CO (dashed) for 40~L.
    }
    \label{fig-3}
\end{figure}

\vspace{2mm} \noindent {\bf Layered ices.}  In Fig.~\ref{fig-3}a TPD
spectra are shown for different layers of $^{18}$O$_2$ on top of 40 L
$^{13}$CO and in Fig.~\ref{fig-3}b (\ref{fig-3}d) 
for $^{18}$O$_2$ ($^{16}$O$_2$) on top of $^{13}$CO ($^{12}$CO)
for different 1/1 configurations (i.e.  20 L $^{18}$O$_2$ /20 L
$^{13}$CO, 40 L $^{16}$O$_2$ / 40 L $^{12}$CO, etc.).  In all cases
the O$_2$ desorption behaves as a 0$^{th}$-order process and is very
similar to that observed for pure O$_2$ ice. This can be expected as
O$_2$ is less volatile than CO, i.e., at the desorbing temperatures of
O$_2$ there is little CO left to influence the O$_2$ desorption
process. The only noticeable difference compared to pure ices is that
the peak temperature shifts to a lower value for O$_2$ and to a higher
temperature for CO by about 0.5~K. This indicates that in the layered
ice systems the binding energies change; for CO the binding energy
increases whereas for O$_2$ it decreases. With exception of the 20~L
$^{13}$CO spectrum, all layered CO traces reveal a 0$^{th}$-order
process as can be seen in Fig.~3b and Fig.~3d for the
$^{12}$CO. Potentially, the presence of O$_2$ on top of the CO ice can
change the desorption process for CO but RAIR spectra of layered ices
do not change much with respect to pure ices. The lack of
co-desorption of CO with the O$_2$ suggests that the molecular
interaction between these species is weak.

\vspace{2mm} \noindent {\bf Mixed ices.}  In Fig. 3c the TPD spectra
are shown for different layer thicknesses of 1:1
$^{18}$O$_2$:$^{13}$CO mixed ices. The O$_2$ desorption follows again
a 0$^{th}$-order process, but the band as a whole is slightly
broadened by 10-15\% compared to pure O$_2$. In the $^{13}$CO spectra
a shoulder around 29.6~K appears, i.e. at the O$_2$ desorption
temperature, which suggests that a fraction of the CO desorbs from
CO-CO binding sites like in pure ices whereas the rest desorbs from a
mixed environment, e.g. CO-O$_2$ binding sites. The main isotopes (not
shown in the figure) exhibit a similar behavior.  Compared with
previous experiments performed on CO-N$_2$ ices (\"Oberg et al. 2005,
Bisschop et al. 2006) these results have been interpreted as follows
(Fuchs et al. 2006):  neither N$_2$ nor O$_2$ possess an
electric dipole moment so they interact with CO mainly via quadrupole
interactions.  However, solid O$_2$ has a 4 to 6 times weaker
quadrupole moment compared to N$_2$ and CO.  Furthermore, N$_2$ and CO
possess the same $\alpha$-crystalline structure below 30~K, but
$\alpha$-phase O$_2$ has a different crystalline structure and also
undergoes a phase change to the $\beta$-form at 23.5~K.  The
combination of these two effects can lead to the absence of mixing and
co-desorption in the CO-O$_2$ system compared with CO-N$_2$, as
observed in our experiments.

\begin{figure}[t]
      \includegraphics[angle=270,width=9cm]{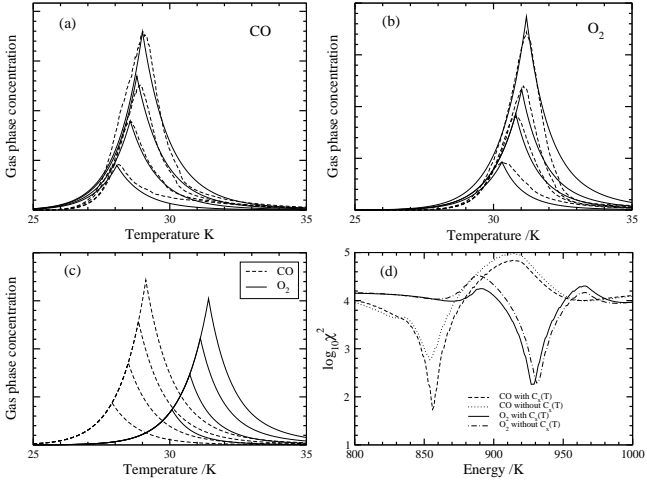}
     \caption{Model TPD spectra of pure ice (solid lines) compared with laboratory
data (dashed lines) for 20, 40, 60 and 80~L ices.
    (a) $^{13}$CO, (b) $^{18}$O$_2$, (c) Model $^{13}$CO (dashed) and $^{18}$O$_2$
(solid) using recommended energy values and 
    (d) variation of $\chi^2$ with energy for an exposure of 40 L, illustrating the well-determined minima.
        The effect of fits with and without using the cryo-pumping function $C_x(T)$ are shown.}
 \label{mod-1-fig}
\end{figure}

\begin{figure}[h]
\centering
    \includegraphics[angle=270,width=9cm]{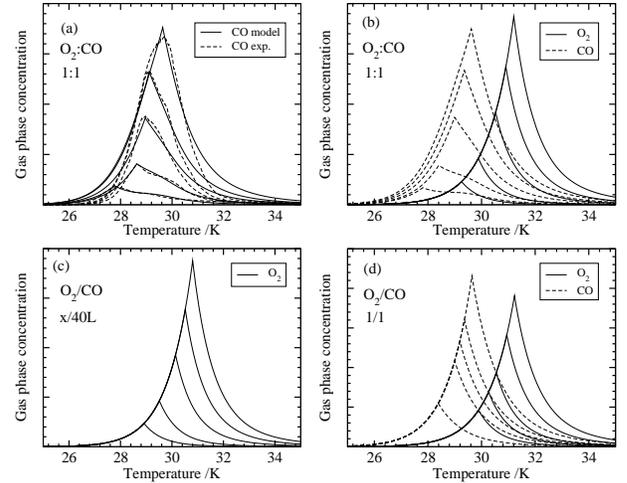}
    \caption{Model TPD spectra of layered and mixed ices; (a) Experimental (dashed)
and model (solid) results of $^{13}$CO 
    in mixed ice; (b) Best fitting $^{13}$CO and $^{18}$O$_2$ model results for
mixed ices; (c) Best fitting
    $^{18}$O$_2$ model results for x/40 L $^{18}$O$_2$ / $^{13}$CO and (d) best
fitting model results for $^{13}$CO and 
    $^{18}$O$_2$ in 1/1 layered ices.}
\label{mod-2-fig}
\end{figure}

\section{Empirical kinetic model of CO-O$_2$ desorption}
The experimental TPD results are interpreted in terms of an empirical kinetic model,
describing the desorption kinetics and providing values for fundamental molecular 
properties to be used in astrochemical models.

\subsection{The model}
The kinetic desorption process for a species $X$ 
can be expressed  by the well known
Polanyi-Wigner type equations of the form,

    \begin{equation}
    R_{des}=\frac{dN_g(X)}{dt}=\nu_i\left[ N_s(X)\right]^i 
\, {\exp [- \frac{E_{d}(X)}{T}]}
    \end{equation}

\noindent
with $R_{des}$ the desorption rate (molecules cm$^{-2}$s$^{-1}$), $N_g(X)$
the number density (cm$^{-2}$) of molecules  desorbing from the substrate, $\nu_i$ a
pre-exponential factor (molecules$^{1-i}$ cm$^{2(i-1)}$s$^{-1}$) for
desorption order $i$, $N_s(X)$ the number density (cm$^{-2}$) of molecules  
on the surface at
a given time $t$, 
$E_{d}(X)$ the binding energy (in K) and $T$ the surface temperature 
(also in K).
The desorption order reflects the nature of the desorption process and
is expressed in integer values although also non-integer values are
allowed.  

Ideally, the pre-exponential factor and the desorption energy are
determined independently.  
However, since both
parameters have a similar effect on the fitting routine a unique
parameter set could not be found. Therefore, the pre-exponential
factor is approximated by the harmonic oscillator of a solid in terms
of a vibrational frequency (s$^{-1}$) by
     \begin{equation}
     \nu=\sqrt{\frac{2 \, N_s \,E_{d}(X)}{\pi^2 \,M}}
     \end{equation}
\noindent
with $N_s$ $\approx$ $10^{15}$ cm$^{-2}$ and $M$ the
mass of species $X$ (Hasegawa et al. 1992).
This equation gives values around 10$^{12}$ s$^{-1}$.  In previous
papers (see e.g., Bisschop et al. 2006, Collings et al. 2003) the
pre-exponential factor has been taken as a free parameter.  In the
present work this parameter has been linked to $E_{d}(X)$ following
Eq.(2).  For a 0$^{th}$-order desorption process, $\nu$ has been
multiplied with the surface density $N_s$, i.e. $\nu_0 = \nu \cdot N_s
= \nu \cdot N_s(t=0)$.  For a 1$^{st}$-order process the vibrational
frequency is taken as a pre-exponential factor, i.e. $\nu_1 =
\nu$. Thus the only parameter that is floating in the present model 
fit to the experimental data is the binding energy $E_{d}(X)$ of the
species.

\begin{table*}[t]
\caption{Best fitting model parameters (recommended values) for CO and O$_2$.}
\label{para-table}
\centering
\begin{tabular}{c c c c c c}
\hline\hline
 Type of ice & Reaction         & Rate equation   & $\nu$                       & $E_{d}$  & $i$ \\
             &                  &                 & [molecules $^{1-i}$         &          &   \\
             &                  &                 & cm$^{2(i-1)}$s$^{-1}$]      &  [K]     &   \\
\hline 
\multicolumn{2}{l}{$^{12}$CO-$^{16}$O$_2$} \\
 Pure    & CO(s)    $\rightarrow$ CO(g)       & $\nu_0 \,\exp(-E_d/T)$     & 7.2E26  &858$\pm$15 & 0 \\
 Pure    & O$_2$(s) $\rightarrow$ O$_2$(g)    & $\nu_0 \,\exp(-E_d/T)$     & 6.9E26  &912$\pm$15 & 0 \\
 Layered & CO(s)    $\rightarrow$ CO(g)       & $\nu_0 \,\exp(-E_d/T)$ & 7.2E26  &856$\pm$15 & 0 \\
 Layered & O$_2$(s) $\rightarrow$ O$_2$(g)    & $\nu_0  \,\exp(-E_d/T)$    & 6.9E26  &904$\pm$15 & 0 \\
 Mixed   & CO(s)    $\rightarrow$ CO(g)       & $\nu_1[{\rm CO}] \,\exp(-E_d/T)$ & 7.6E11  &955$\pm$18 & 1 \\
         & CO(s)    $\rightarrow$ CO(g)       & $\nu_0 \, \exp(-E_d/T)$    & 7.2E26  &865$\pm$18 & 0 \\
 Mixed   & O$_2$(s) $\rightarrow$ O$_2$(g)    & $\nu_0 \, \exp(-E_d/T)$    & 6.8E26  &896$\pm$18 & 0 \\
\\
\multicolumn{2}{l}{$^{13}$CO-$^{18}$O$_2$} \\
Pure     & CO(s)    $\rightarrow$ CO(g)       & $\nu_0 \,\exp(-E_d/T)$     &7.0E26  & 854$\pm$10 & 0 \\
Pure     & O$_2$(s) $\rightarrow$ O$_2$(g)    & $\nu_0 \,\exp(-E_d/T)$     &6.6E26  & 925$\pm$10 & 0 \\
Layered  & CO(s)    $\rightarrow$ CO(g)       & $\nu_0 \,\exp(-E_d/T)$ &7.1E26  & 860$\pm$15$^*$ & 0 \\
Layered  & O$_2$(s) $\rightarrow$ O$_2$(g)    & $\nu_0  \,\exp(-E_d/T)$    &6.5E26  & 915$\pm$10 & 0 \\
Mixed    & CO(s)    $\rightarrow$ CO(g)       & $\nu_1[{\rm CO}] \,\exp(-E_d/T)$ &7.5E11  & 965$\pm$10 & 1 \\
         & CO(s)    $\rightarrow$ CO(g)       & $\nu_0 \, \exp(-E_d/T)$    &7.2E26  & 890$\pm$10 & 0 \\
Mixed    & O$_2$(s) $\rightarrow$ O$_2$(g)    & $\nu_0 \, \exp(-E_d/T)$    &6.2E26  & 915$\pm$10 & 0 \\
\\
Pump \\
         & CO(g)    $\rightarrow$ CO(pump)    & k$_{pump}$[CO(g)]          & 0.00024 & - & 1 \\   
         & O$_2$(g) $\rightarrow$ O$_2$(pump) & k$_{pump}$[O$_2$(g)]       & 0.00036 & - & 1 \\
\hline
\end{tabular}
\\
$^*$ Average value for ice thicknesses greater than 30~L. For ices between 20 and 30~L use 888$\pm$15~K.
\end{table*}

\vspace{2mm} \noindent
Starting from Eq.~(1), the kinetics are represented in terms of two coupled
differential equations

     \begin{equation}
      \frac{dN_{s}(X)}{dt} = - \sum_{i=0,1} \nu_{i} \left[ N_s(X) \right]^i {\exp [- \frac{E_{d,i}(X)}{T}]}
     \end{equation}

\pagebreak

     \begin{equation*}
     \frac{dN_{g}(X)}{dt} =  \sum_{i=0,1} \nu_{i} \left[ N_s(X) \right]^i {\exp [- \frac{E_{d,i}(X)}{T}]} 
     \end{equation*}
     \begin{equation}
      - C_x(T)k_{pump}N_{g}(X)
     \label{eq-des}
     \end{equation}

\noindent
representing the density change of the solid phase (s)  and
gas phase (g) molecules.  The last term of Eq.~(\ref{eq-des}) gives
the removal of gaseous species from the vacuum chamber by the
pump. Here, $k_{pump}$ is the pumping constant in s$^{-1}$ and
$C_x(T)$ is a dimensionless cryo-pumping factor between 1 and 7 as
discussed in \S\ref{exp-sec}.  The pumping constant, $k_{pump}$, is
constrained by fitting the pump down curve of both species, i.e., the
slope of the TPD curve at temperatures higher than the desorption peak
temperature.  In most cases, the peak tail is not well reproduced due
to the presence of other cold surfaces in the system. Very small
amounts of gas that have missed the target can get deposited and can
be desorbed afterwards at a temperature different from that of the
target surface.  A direct consequence is a deviation in model curves
from the experimental plots at high temperatures. However, this does
not affect the determination of the binding energies that are mainly
sensitive to the peak value and the leading edge.

In order to calculate the temperature dependent rate measured in the TPD
experiments, $dn/dt$ is written as

     \begin{equation}
     \frac{dn}{dt}=\frac{dn}{dT}\frac{dT}{dt}=\beta\frac{dn}{dT}
     \end{equation}

\noindent
with $dn/dT$ the temperature dependent rate (molecules
cm$^{-2}$K$^{-1}$), and $\beta$ = $dT/dt$ the constant TPD heating
rate (K s$^{-1}$) that corresponds to 0.1 K/minute in the present
study unless stated differently.

\vspace{2mm} \noindent To extract the binding energy from the
observations a standard minimization of $\chi^2$ is used which
represents the sum over the squares of the differences between the
experimental points and the calculated ones. In this procedure first a
0$^{th}$- or 1$^{st}$-order process is fitted and subsequently the
appropriate parameters are optimized.

\subsection{Results}
The results are shown in Figs.~\ref{mod-1-fig} and \ref{mod-2-fig} and Table 2 summarizes the model equations 
with best fit parameters. 
In nearly all experiments a linear dependence between ice thicknesses and desorption energies has been 
found.

\vspace{2mm} \noindent {\bf Pure ices.}  For pure $^{18}$O$_2$ five
experiments are performed using different thicknesses (20 L, 40 L, 50
L, 60 L, 80 L ) and the best fits to these experiments
(Fig.~\ref{mod-1-fig}b) give 932, 927, 924, 928 and 918 K,
respectively, corresponding to an average value of 925 $\pm$ 10 K. The
error of the mean value is chosen to give a conservative estimate of
these parameters. Similarly, for pure CO (Fig.~\ref{mod-1-fig}a) the
binding energy is 854 $\pm$ 10 K which is consistent with the value of
855~K reported by {\" O}berg et al. (2005) and Bisschop et
al. (2006). Both pure CO and pure O$_2$ ice exhibit 0$^{th}$-order
kinetics.  The TPD spectra are reproduced very well with this model
using only one free parameter as is demonstrated in
Figs.~\ref{mod-1-fig}a and b, where the experimental and model results
are plotted on top of each other. In Fig.~\ref{mod-1-fig}d the
variations of $\chi^2$ for the experiments of 40 L for the pure CO and
O$_2$ are shown. The energy values for which $\chi^2$ is minimum are
taken as the final values. This yields 854~K for $^{13}$CO and 927~K
for $^{18}$O$_2$.  As mentioned in the experimental section we have
explicitly taken into account the effect of cryo-pumping.  A
systematic deviation of 2 to 3~K in the desorption energy is induced
when this factor is neglected.  This is illustrated in
Fig.~\ref{mod-1-fig}d.  Inclusion of the $C_x(T)$ function generally
improves the fit substantially.\\ $^{12}$CO and $^{16}$O$_2$ show a
similar desorption behavior and the binding energies are 858 $\pm$
15~K and 912 $\pm$ 15~K, respectively. Thus $^{12}$CO has the same
binding energy as $^{13}$CO within the errors.  For $^{16}$O$_2$ ices
the binding energy is lowered by 1.6\% with respect to $^{18}$O$_2$ on
average. Because of the smaller data set for the main isotope
the errors in the fitted parameters of the $^{12}$CO and $^{16}$O$_2$ isotopes are
larger than those for $^{13}$CO and $^{18}$O$_2$.

\vspace{2mm} \noindent {\bf Layered ices.}  The desorption of
$^{18}$O$_2$ in 1/1 (Fig.~\ref{mod-2-fig}d) and x/40
(Fig.~\ref{mod-2-fig}c) layered ices is 0$^{th}$-order and the binding
energy is determined as 915 $\pm$ 10 K, which is slightly lower than
for the pure ices. $^{16}$O$_2$ shows a similar behavior with a
binding energy of 904 $\pm$ 10 K. $^{13}$CO desorption is a
0$^{th}$-order process and only at low coverages a 1$^{st}$-order
process may be involved. Since there is no signature in the TPD
spectra (nor in the RAIR spectra) of mixing or segregation and since
the spectrum looks very similar to pure ices it is modeled in
0$^{th}$-order with an average binding energy of 860 $\pm$ 15 K for
ices thicker than 30~L.  The correctness of this procedure is
confirmed by the TPD spectra of $^{12}$CO which exhibit a
0$^{th}$-order desorption with a binding energy of 856 $\pm$ 15 K.

\vspace{2mm} \noindent {\bf Mixed ices.}  O$_2$ in mixed ices
(Fig.~\ref{mod-2-fig}c) is again 0$^{th}$-order with a similar binding
energy as in layered ices. In mixed ices, TPD spectra of $^{13}$CO are
broader with respect to pure ices and a shoulder around the desorption
temperature of O$_2$ is observed.  This suggests that CO is desorbing
from a wider range of binding sites and this is also supported by RAIR
spectra that are broader with respect to pure ice spectra.  Both the
peak and shoulder are fitted very well using a combination of
0$^{th}$- and 1$^{st}$-order processes (see Fig.~\ref{mod-2-fig}a).
About 50\% of the CO desorbs from CO-CO binding sites as in the pure
ices but with a binding energy of 890 $\pm$ 10 K. The residual CO
molecules desorb through a 1$^{st}$-order process from a mixed
environment, i.e. including CO-O$_2$ binding sites, with a binding
energy of 965 $\pm$ 5~K.

Table 2 has separate entries for the $^{13}$CO-$^{18}$O$_2$ and
$^{12}$CO-$^{16}$O$_2$ isotopomers.  Independent of the used
isotopomers it can be concluded that the binding energies of CO
increase in the following order $E_{\rm pure}$ $\lesssim$ $E_{\rm
layered}$ $<$ $E_{\rm mixed}$. For O$_2$ this is inverted $E_{\rm
pure}$ $>$ $E_{\rm layered}$ $\gtrsim$ $E_{\rm mix}$.

\vspace{2mm} \noindent {\bf The effect of $\beta$ on the binding
energies.}  In order to rule out any dependencies on the adopted heating
rates, experiments have been performed for $\beta$=0.1, 0.2 and 0.5
K min$^{-1}$ on 40~L CO ices.  This is of relevance when applying laboratory
values to interstellar warm-up time scales.  The calculated $\nu$ and
$E_{d}$ values for the 0.5 K min$^{-1}$ experiment have been used to predict
the experimental TPD curve for 0.2 and 0.1 K min$^{-1}$.  The deviation
between the calculated and experimental values are within the
experimental error, i.e. $\Delta E_{d}$ = $\pm$ 15~K, with a slight
tendency for lower inferred $E_{d}$ at lower
heating rates.

\subsection{Sticking probability}
\begin{figure}[t]
    \includegraphics[angle=270,width=9cm]{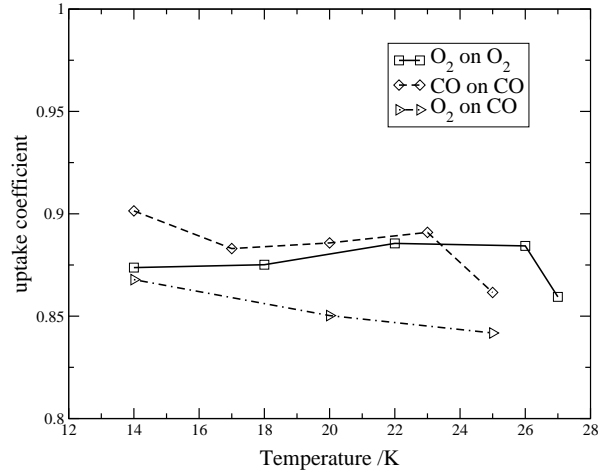}
    \caption{Uptake coefficient $\gamma$ representing the lower limits 
on the sticking coefficients as functions of the
temperature below 25~K. }
\label{stick-fig}
\end{figure}

In addition to the desorption rates, the sticking coefficients also
play an important role to describe freeze-out onto interstellar
grains. The measurement procedure has been extensively discussed in
Bisschop et al. (2006) and Fuchs et al. (2006). It is important to
note that the measurements only provide an `uptake' coefficient
$\gamma$ rather than a sticking coefficient, $S$, as only the net rate
of molecules sticking and leaving the surface can be
given. Consequently, the values given in Fig.~\ref{stick-fig}
represent lower limits for the sticking coefficients of O$_2$ and
CO. It is found that the freeze-out dominates for O$_2$ on O$_2$, CO
on CO and O$_2$ on CO up to 25~K with lower limits on the sticking
probabilities between 0.85 and 0.9, i.e., close to unity. Under real
astrophysical conditions this value will increase and approaches unity
at 10 K.

\section{Astrophysical implications}

The scenarios put forward to explain the absence of gaseous O$_2$ in
interstellar clouds can roughly be divided into two categories:
time-dependent models of cold cores invoking freeze-out of oxygen (in
all its forms) onto grains (e.g., Bergin et al.\ 2000, Aikawa et al.\
2005), and depth-dependent models of large-scale warm clouds invoking
deep penetration by UV radiation in a clumpy structure enhancing the
O$_2$ photodissociation (e.g., Spaans \& van Dishoeck 2002). The
laboratory experiments presented here are relevant to the first
scenario.  In these models, atomic O is gradually transformed with
time into O$_2$ in the gas phase. Freeze-out of oxygen occurs on a
similar timescale, and any O is assumed to be turned effectively into
H$_2$O ice on the grains where it subsequently sticks. No
formation of solid O$_2$ on the grain is expected because of the
presence of atomic hydrogen, which is much more mobile at low
temperatures. The H$_2$O ice formation lowers the gaseous [O]/[C]
ratio and potentially leaves only a small abundance of O$_2$ in the
gas. The amount of solid O$_2$ and the remaining fraction of gaseous
O$_2$ of $\sim 10^{-8} - 10^{-7}$ with respect to H$_2$ depends
sensitively on the temperature at which O$_2$ is frozen out. A related
question is to what extent O$_2$ differs in this respect from CO,
since CO is readily observed in the gas phase and in solid form.

\begin{figure*}[t]
     \includegraphics[width=17cm]{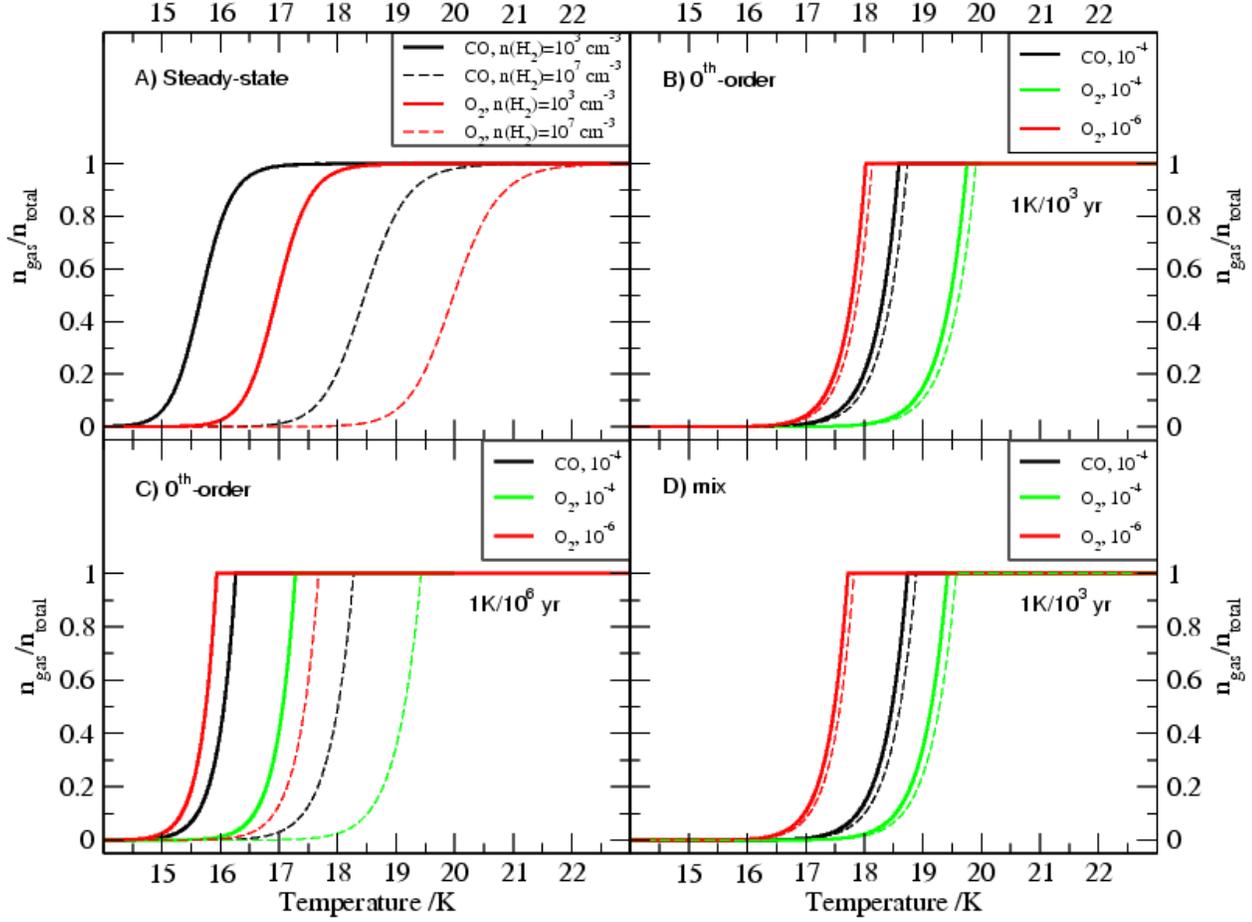}
\caption{Astrophysical simulations of gaseous CO and O$_2$
         abundances relative to the total gas + solid abundances. Panel (a)
         shows pure CO and O$_2$ in steady-state, panel (b) 0$^{th}$-order
         desorption for heating rates of 1~K/1000 yr, panel (c) the same but
         for 1~K/10$^6$ yr, and panel (d) shows the desorption of CO from mixed
         (layered) ices for 1~K/10$^3$ yr.  In the panels (b)-(d)
         the solid line indicates accretion corresponding to 
         $n$(H$_2$)=10$^3$ cm$^{-3}$, the dashed line corresponds to
         accretion for $n$(H$_2$)=10$^7$ cm$^{-3}$. 
         For the panels (b)--(d) the adopted initial gaseous abundances of
         CO is $10^{-4}$  and for O$_2$ it is $10^{-4}$ and $10^{-6}$ with respect to H$_2$.
}
\label{astro-fig}
\end{figure*}

Astrochemical models usually assume 1$^{st}$-order desorption with
rates in cm$^{-3}$ s$^{-1}$. Appendix A  (see online material)
summarizes the equations used
to apply our laboratory results to astrochemical models for both
1$^{st}$ and 0$^{th}$ order kinetics. The latter is more appropriate
for thick ices in interstellar clouds and consistent with our
laboratory data.  We consider the simple case of pure desorption and
accretion of $^{12}$CO and $^{16}$O$_2$, without any gas-phase or
solid-state chemistry.  Thus, the main difference with the equations
in \S 4.1 is that desorption is now balanced by a freeze-out
(accretion) term, rather than a pumping factor.  Also, the laboratory
heating rate can be replaced by a heating rate appropriate for the
astrophysical object under consideration, for example a protostar
heating its envelope.

Our model only considers classical dust grains with a radius of
0.1~$\mu$m. The density $n$(H$_2$) is varied to simulate the effects
for clouds of different densities.  In 0$^{th}$ order, the results
depend on the initial number densities of the CO and O$_2$ molecules.
For CO, an abundance $10^{-4}$ relative to H$_2$ is chosen, initially
all on the grains. For O$_2$, two options are considered. The first
option is that most oxygen has been converted into O$_2$ at an
abundance of $10^{-4}$, as found in gas-phase models and the maximum
allowed by the observed upper limits on solid O$_2$.  The second case
considers a much lower O$_2$ abundance of 10$^{-6}$, as found in
gas-grain models.

Fig.~\ref{astro-fig}a shows the results for pure CO and pure O$_2$ ices
in the simplest case of steady-state. The temperature at which these
species desorb depends strongly on density: the higher the density,
the faster the accretion rate which needs to be balanced by
desorption. Because of the larger binding energy of O$_2$, it always
desorbs at $\sim$2 K higher temperatures than CO.

Fig.~\ref{astro-fig}b shows the result for 0$^{th}$ order desorption
at a heating rate typical for a proto-stellar environment of 1 K/1000
yr (Lee et al. 2005).  
For abundances of CO and O$_2$ of $10^{-4}$, both species desorb at
$\sim$2.5 K higher temperatures than the steady-state case at low
densities; at high densities, there is little difference.  If the
O$_2$ abundance is lowered to $10^{-6}$ the 0$^{th}$ order curve
shifts to lower temperature by $\sim$2 K. Thus, in contrast
with the steady-state or equal abundance cases, O$_2$ can desorb at
{\it lower} temperatures than CO if its abundance is significantly
lower. This reversal is a specific feature of the 0$^{th}$ order
desorption and is not found in the 1$^{st}$ order formulation (see
Appendix A in the online version). 
An important question is whether such a situation is
astrochemically relevant. For O$_2$ abundances as low as $10^{-6}$,
the coverage becomes less than a mono-layer and 1$^{st}$ order
kinetics or other effects due to the peculiarities of the ice 
(e.g. polar vs. apolar environment, compact vs. porous ice)
will determine the desorption behavior.

Fig.~\ref{astro-fig}c shows the abundance curve of the
species for a heating rate of 1~K/10$^6$ yr, as expected for a cold
core at near constant temperature.  Since the time scale is increased
by 3 orders of magnitude with respect to the previously considered
heating rate there is simply more time for the molecules to desorb and
consequently the entire profile shifts to lower temperatures by
0.5 -- 2.5 K depending on the density of the species.

Finally, Fig.~\ref{astro-fig}d shows the abundance curves of CO and O$_2$ from
mixed ices.
The graphs of mixed and layered ices are nearly identical, thus 
the layered ices are not shown separately.
The relative abundance curve is shifted to a
higher temperature by about 0.5~K compared with the pure ices because
of the slightly higher binding energies.

Overall, Fig.~\ref{astro-fig} shows that the differences in the
desorption behavior of O$_2$ and CO with temperature are very minor
for a wide range of realistic cloud densities and abundances.  Thus,
it is unlikely that a large reservoir of solid O$_2$ is hidden in the
bulk of molecular clouds which show abundant gaseous CO but no O$_2$,
unless the O$_2$ is in a more strongly bound ice
environment. Conversely, any region with significant CO freeze-out
should also have some solid O$_2$.  As noted in the introduction,
the best limits come from analysis of the weak solid $^{13}$CO band,
which gives upper limits of 100\% on the amount of O$_2$ that can be
mixed with CO, i.e., about $(0.5-1)\times 10^{-4}$ with respect to
H$_2$ (Boogert et al.\ 2003, Pontoppidan et al.\ 2003). Direct
freeze-out of the gas-phase O$_2$ abundances inferred from the ODIN
measurements would give much lower limits.

The small differences between CO
and O$_2$ desorption found here may become relevant in the
interpretation of high spatial resolution observations of individual
cold cores with temperatures in the 10--20 K range such as the pre-stellar
core B68.

\section{Concluding remarks}
The desorption processes of CO-O$_2$ pure, mixed and layered ice
systems have been investigated experimentally and modeled using an
empirical kinetic model. The resulting molecular parameters can be
used to model the desorption behavior of these ices under
astrophysical conditions.  We find that both pure $^{16}$O$_2$ and
pure $^{12}$CO desorb through 0$^{th}$-order processes with binding
energies of 912 $\pm$ 15 K and 858 $\pm$ 15 K, respectively.  In mixed
and layered ices the $^{16}$O$_2$ binding energy decreases to a lower
value around 896 $\pm$ 18 K and 904 $\pm$ 15~K respectively.  The
$^{12}$CO desorption from layered ices is 0$^{th}$-order with a
binding energy of 856 $\pm$ 15~K. In mixed ices a combination of
0$^{th}$- and 1$^{st}$-order is found with desorption energies of 865
$\pm$ 18 K and 955 $\pm$ 18 K, respectively. For $^{18}$O$_2$ and
$^{13}$CO, these numbers change by a few percent.

O$_2$ is less volatile than CO but CO does not co-desorb with O$_2$.
This is in contrast with the CO--N$_2$ ice system for which Bisschop
et al. (2006) found that N$_2$ is more volatile than CO and that
significant amounts of N$_2$ co-desorb with CO. The sticking
coefficients of CO and O$_2$ at temperatures below 20~K are close to
unity, with 0.85 as a lower limit.  In cold clouds ($T_d<18$~K),
O$_2$ can be frozen out onto the grains, but the relative difference
in desorption between CO and O$_2$ is so small that this is unlikely
to be the explanation for the missing gaseous O$_2$ in interstellar
clouds which show significant gaseous CO.

\begin{acknowledgements}
We are grateful to F.A. van Broekhuizen, S.E Bisschop, K.I. {\"O}berg
and S.  Schlemmer for useful discussions and help in the construction
of the experimental setup. We acknowledge funding through the
Netherlands Research School for Astronomy (NOVA), FOM and a Spinoza
Grant from the Netherlands Organization for Scientific Research
(NWO). K.A. thanks the Greenberg family for a Greenberg research
fellowship and the ICSC-World Laboratory Fund for additional funding.
\end{acknowledgements}

{}

\Online

\appendix
\section{Modeling accretion and desorption in astrochemical applications}

The gas-phase density $n_g(X)$ (in cm$^{-3}$) of species $X$ with respect to its
total gas + solid density $n_{\rm tot}(X) = n_g(X) + n_s(X)$ can be
calculated from the accretion and desorption rates. 
For reasons of simplicity and to compare with other work
we consider here
the simplest case of a single type of grain with a classical radius
$r_c$=0.1 $\mu$m, $10^{15}$ sites per cm$^2$ and a grain abundance
$n_{gr}$/$n$(H$_2$)=$10^{-12}$.

\subsection{Accretion rate}

The most general formulation of the accretion rate in  cm$^{-3}$ s$^{-1}$ is
\begin{equation}
R_{acc} = \sigma S v n_{g}(X)  n_{gr} 
\label{acc-1}
\end{equation}
where $\sigma$=$\pi r_c^2$ is the grain cross
section, $S$ is the sticking coefficient (taken to be unity at
$T_{gr}<20$~K) and $v$ is the mean speed of the gas in cm s$^{-1}$.
For a classical grain and the canonical grain abundance this becomes
\begin{equation} 
R_{acc} = 4.55 \times 10^{-18} \biggr( \frac{T}{M(X)} \biggr)^{0.5} 
   n_g(X) n_{H_2}   
\label{acc-2}
\end{equation}
where $T$ is the gas temperature and $M(X)$ is the mass of $X$ in
atomic mass units (amu).
The accretion rate per molecule $X$ in s$^{-1}$ is denoted as
\begin{equation} 
\lambda =   R_{acc} / n_g(X) 
\label{acc-3}
\end{equation}
and the timescale for freeze-out can be computed from
$\tau=1/\lambda$. Note that in dense cold cores, the grains may
have grown to larger $\mu$m sizes than assumed here due to
coagulation, in which case the abundance has to be lowered. Also, a
more sophisticated calculation should include a grain size
distribution.

\subsection{Desorption rate}

The desorption rate in the laboratory surface science experiments is
given by Eq.\ (1) in cm$^{-2}$ s$^{-1}$, which is rewritten here as
\begin{equation}
R_{des,lab} = \nu_i \, N^i \, \exp \biggr[- \frac{E_i}{kT_{gr}} \biggr]
\label{des-1}
\end{equation}
where $N$ has the units of molecules cm$^{-2}$, and the unit of $\nu_i$
changes with the order $i$ of desorption as molecules$^{1-i}$ \,
cm$^{2(i-1)}$ \, s$^{-1}$.  For 0$^{th}$-order $\nu_0$ is in
molecules cm$^{-2}$ s$^{-1}$ and for 1$^{st}$-order $\nu_1$ is in
s$^{-1}$. $E_i$ is the binding energy for order $i$.

\subsubsection{1$^{st}$ order}

For 1$^{st}$ order kinetics, the translation from the laboratory data to a
astronomical desorption rate $\xi$ in s$^{-1}$ is straight-forward:
\begin{equation}
\xi_{1} = \nu_1  \exp \biggr[- \frac{E_1}{kT_{gr}} \biggr]
\label{des-4}
\end{equation}
with $\nu_1$ and $E_1$ taken directly from the laboratory data. The
rate $R_{des}$ in cm$^{-3}$ s$^{-1}$ becomes
\begin{equation}
R_{des} = \xi_1 n_s(X)= \nu_1 n_s(X) \exp \biggr[- \frac{E_1}{kT_{gr}} \biggr]
\label{des-5}
\end{equation}
with $n_s$ (in cm$^{-3}$) the density of molecules on the grain surface.

\subsubsection{0$^{th}$ order}

For 0$^{th}$ order kinetics, the desorption rate $R_{des}$ in
cm$^{-3}$ s$^{-1}$ should be proportional to the number
density of grains $n_{gr}$ and the average surface area $A_{gr}$ of
a grain.  The rate `per grain' in s$^{-1}$ can be obtained from

\begin{equation}
\xi_{0} = \nu_0 A_{gr} \exp(-E_0/kT_{gr}) 
\label{des-2}
\end{equation}
with $\nu_0$ and $E_0$ derived from the laboratory data for 0$^{th}$ order.
The total rate is 
\begin{equation}
    \begin{split}
      R_{des}  & = \xi_{0} n_{gr} = \nu_0 A_{gr} n_{gr}\exp(-E_0/kT_{gr}) \\ 
    \end{split}
\label{des-3}
\end{equation}

\subsection{Steady-state model} 

If no chemical reactions in the gas or on the grains are included,
a steady-state will be reached where the accretion balances the
desorption. In 1$^{st}$ order, this becomes
\begin{equation}
     \lambda  n_g(X)  = \xi_1 n_s(X) 
\label{bal-1}
\end{equation}
with  $\xi_1$ from Eq.~\ref{des-4}.
Thus the ratio of gas-phase to solid-state molecules is given by 
\begin{equation}
    \frac{n_g}{n_s} = \frac{\xi_1}{\lambda} = 2.2 \cdot 10^{17} {\nu}_1 \, \sqrt{M} 
  \quad \frac{\exp(-E_1/kT_{gr})}{\sqrt{T}}
\label{bal-2}
\end{equation}

Results of Eq.\ref{bal-2} are shown in Fig.~\ref{fig-A1} for two
densities $n$(H$_2$)=$10^3$ and $10^7$ cm$^{-3}$.  No heating rate
$\beta$ is involved in this calculation.  It is the most simple
``astrochemical'' modeling result achievable.  One note of caution,
however: it is not correct to use laboratory values for $E_i$ which
were fitted to 0$^{th}$-order laboratory data for 1$^{st}$-order
astrochemical equations and just change $\nu$. If 1$^{st}$ order
kinetics are used (even though the laboratory data indicate 0$^{th}$
order), a better approach would be to first fit the laboratory data
with 1$^{st}$ order kinetics and then apply those in the astrochemical
models. The differences in binding energies can be as large as 12\%
for CO and O$_2$. For pure N$_2$, CO and O$_2$, 1$^{st}$ order fits to
the peaks of our laboratory curves give 800 K, 955 K and 1035 K,
respectively, with $\nu_1=1.0 \cdot 10^{11}$ s$^{-1}$ for N$_2$
(Bisschop et al.\ 2006) and $\nu_1=1.0 \cdot 10^{12}$ s$^{-1}$ for CO and O$_2$.
Figures~\ref{astro-fig}a and \ref{fig-A1} use these values.

\subsection{Time-dependent model}

\subsubsection{1$^{st}$ order}

In the 1$^{st}$ order approach, the desorption rate changes with the
number of species $X$ bound to the grain, which is justified only for
(sub)monolayer or irregular coverage.  In this scenario the abundance
of a species is determined by two differential equations and some
boundary (initial) conditions

\begin{equation}
    \begin{split}
          dn_g/dt = - \lambda n_g + \xi_1 n_s \\
          dn_s/dt = + \lambda n_g - \xi_1 n_s. \\
    \end{split}
\label{tev-1}
\end{equation}

These are the equations commonly used in time-dependent astrochemical
models (e.g., Bergin \& Langer 1997, Charnley et al.\ 2001, Aikawa et
al.\ 2005).

For protostellar cores in which a star turns on, the simplest
assumption is that there is a linear relation between the temperature
$T_{gr}$ and the time $t$
\begin{equation}
    T_{gr}(t) = T_{gr,0} + dT_{gr}/dt \cdot t = T_{gr,0} + \beta \, t
\label{tev-2}
\end{equation}
with $T_{gr,0}$ around 10~K. At $t=0$, all molecules are assumed to be
frozen out onto grains, $n_s(t=0)=n_{\rm tot}(X)$, as appropriate for
a pre-stellar core prior to star formation. A completely opposite
situation, not studied here, would be a quiescent cloud where
$dT_{gr}/dt$ is close to zero and all molecules are initially in the
gas phase. Note that in this case the timescales are dominated by the
freeze-out process rather than the evaporation timescale, and that
this results in a very different time behavior, depending on
temperature.

Fig.~\ref{fig-A1} includes the time-dependent 1$^{st}$ order results
for a heating rate of 1 K per $10^3$ yr and two densities. Compared
with the steady-state results, the low density curves are shifted to
higher temperatures by $\sim$3 K, whereas the high density curves are
nearly identical. A slower heating rate of 1 K per $10^6$ yr 
brings the low density curves closer to the low density steady-state results.

\subsubsection{0$^{th}$ order}

In the 0$^{th}$ order approach, the desorption rate is constant with
time, independent of the surface density $n_s(X)$. The time-dependent
equations now become

\begin{equation}
    \begin{split}
          dn_g/dt   &= - \lambda n_g +  R_{des} \quad 
   \text{for {\it n$_s$} $>$ 0} \\
                      &= - \lambda n_g             \hspace{1.6cm} 
   \text{for {\it n$_s$} = 0}   \\
          dn_s/dt   &= + \lambda n_g -  R_{des}  \quad  
     \text{for {\it n$_s$} $>$ 0} \\
                      &= + \lambda n_g             \hspace{1.6cm}   
 \text{for {\it n$_s$} = 0}   \\
    \end{split}
\label{tev-5}
\end{equation}
with $R_{des}$ from Eq.~\ref{des-3}. This formulation is most
appropriate for thick, fully covered ice layers. For partial coverage,
$R_{des}$ would need to be multiplied by a relative surface occupation
number $\theta$ between 0 and 1, but for the cases of pure CO and
O$_2$ ices considered here $\theta$=1 is taken.  The condition on the
right hand side causes an abrupt end to the desorption process once
the molecules on the surface have been evaporated.  Instead of getting
a smooth $S-$shaped curve like in the 1$^{st}$st-order approach, the
density curve of a 0$^{th}$-order formulation reveals a sharp edge at
the end of the evaporation process.

\begin{figure}[h]
\hspace{-0.7cm}
     \includegraphics[width=10cm]{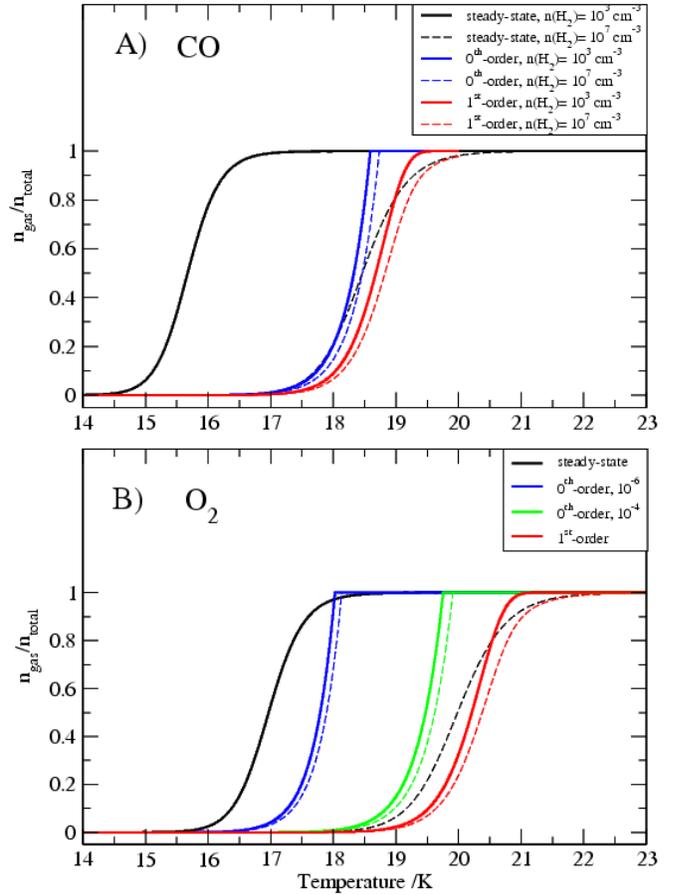}
    \caption{The gas-phase fraction of CO (top) and O$_2$ (bottom) for
densities of $n$(H$_2$)=$10^3$ cm$^{-3}$ (full lines) and $10^7$  cm$^{-3}$ (dashed lines) in the case
of pure ices. Both steady-state (black), 0$^{th}$-order (blue/green) and
$1^{st}$-order (red) desorption are considered. In 0$^{th}$ order, the
adopted abundance for CO is $10^{-4}$ whereas for O$_2$ both $10^{-4}$ (green)
and $10^{-6}$ (blue) are considered. }
\label{fig-A1}
\end{figure}

The results now depend on the choice of the initial $n_g(X)/n$(H$_2$)
abundance ratio of species $X$. For CO, a logical choice is $n_g({\rm
CO})/n({\rm H_2}) = 10^{-4}$. For O$_2$, an extreme case would be to
put all oxygen into O$_2$, $n_g({\rm O_2})/n({\rm H_2}) >
10^{-4}$. However, models and observations (see \S 1) suggest that a
more plausible abundance is at least two orders of magnitude lower.
Figure~\ref{fig-A1} shows the results for $n_g({\rm O_2})/n({\rm H_2})
= 10^{-6}$ and $10^{-4}$.  The 0$^{th}$ order curves for CO and O$_2$
for $10^{-4}$ differ by less than 0.7 K from
those at 1$^{st}$ order. However, those for O$_2$ at $10^{-6}$ are
shifted to lower temperature by $\sim$2.5 K.

These examples illustrate the importance of proper modeling of both
the order of the desorption processes involved and the balance with
accretion. Shifts of a few~K compared to the steady-state solution can
occur either way depending on cloud parameters. Such small shifts are
unlikely to affect the interpretation of large-scale molecular clouds,
but they may potentially become significant in interpretations of the
relative behavior of CO, O$_2$ and N$_2$ in cold dense clouds in the
10--20~K range.

Finally, it should be noted that our results using the empirical
model for pure ices are consistent with first principle considerations
using saturated vapor pressure equations, as done for pure CO by
L\'eger (1983).

\begin{thebibliography}{}

\bibitem[]{}Bergin, E.~A., Langer, W.~D., \& Goldsmith, P.~F. 1995, ApJ, 
441, 222
\bibitem[]{}Bergin, E.~A., Melnick, G.~J., Stauffer, J.~R., et al. 2000,  ApJ, 539, L129
\bibitem[]{}Bisschop, S.~E., Fraser, H.~J., {\"O}berg, K.~I., van Dishoeck, E.~F., Schlemmer, S., 2006, A\&A, 449, 1297
\bibitem[]{}van Broekhuizen, F.~A. 2005, PhD thesis, Leiden University
\bibitem[]{}Chiar, J.~E., Gerakines, P.~A.,  Whittet, D.~C.~B., Pendleton, Y.~J., Tielens, A.~G.~G.~M.,  Adamson, A.~J., \&
 Boogert, A.~C.~A. 1998, ApJ, 498, 716
\bibitem[]{}Collings, M., Dever, J., Fraser, H., \& McCoustra, M. 2003, ApJS,
285, 633
\bibitem[]{}Collings, M.~P., Anderson, M.~A., Chen, R., Dever, J.~W., Viti, S.,
Williams, D.~A., \& McCoustra, M.~R.~S.
2004, MNRAS, 354, 1133
\bibitem[]{}Dartois, E., 2006, A\&A, 445, 959
\bibitem[]{}D'Hendecourt, L.~B., Allamandola, L.~J., \& Greenberg, J.~M. 1985, A\&A, 152, 130 
\bibitem[]{}Ehrenfreund, P., Breukers, R., d'Hendecourt, L., Greenberg, J.M., 1992, A\&A, 260, 431
\bibitem[]{}Ehrenfreund, P., Boogert, A., Gerakines, P., Tielens, A., \& van
Dishoeck, E. 1997, A\&A, 328, 649
\bibitem[]{}Ehrenfreund P. \& van Dishoeck E. 1998, Adv. Space. Res., 21, 15 
\bibitem[]{}Elsila, J., Allamandola, L.~J., \& Sandford, S.~A. 1997, ApJ, 479, 818
\bibitem[]{}Fraser, H.~J., Collings, M.~P., McCoustra, M.~R.~S. and Williams, D.~A.
2001, MNRAS, 327, 1165
\bibitem[]{}Fuchs, G.W., Acharyya, K., Bisschop, S.E., et al., 2006, 
Far. Disc., 133, 331
\bibitem[]{}Goldsmith, P.~F., Melnick, G.~J., Bergin, E.~A., et~al. 2000, 
ApJ, 539, L123
\bibitem[]{}Hasegawa, T.~I., Herbst, E., \& Leung, C.~M. 1992, ApJS, 82,
167
\bibitem[]{}L\'eger, A., 1983, A\&A, 123, 271 
\bibitem[]{}Liseau, R. \& {Odin Team}. 2005, in Astrochemistry: 
recent successes and current challenges, IAU Symposium No 231, 
            eds. Lis, D., Blake, G.A., Herbst, E. 
(Cambridge University Press), p.\ 301
\bibitem[]{}Meyer, D.M., Jura, M, Cardelli, J.A., 1998, ApJ, 493, 222
\bibitem[]{}{\" O}berg, K.~I., van Broekhuizen, F., Fraser, H.~J., 
van Dishoeck, E.~F., Schlemmer, S., 2005, ApJ, 621, L33
\bibitem[]{}Pagani, L., Olofsson, A.~O.~H., Bergman, P., 2003, A\&A, 402, L77
\bibitem[]{}Pontoppidan, K.~M., Fraser, H.~J., Dartois, E., 
et~al. 2003, A\&A, 408, 981
\bibitem[]{}Vandenbussche, B., Ehrenfreund, P., Boogert, A.~C.~A., 
et al. 1999, A\&A, 346, L57
\end{thebibliography}
\end{document}